\documentclass{aa} \usepackage{graphics} 
\begin{document} 
\thesaurus{11.05.2,11.06.2,11.09.5,11.19.2,13.21.1} 
\title
{The ultraviolet visibility and quantitative morphology of galactic disks 
at low and high redshift}

\author {D. Burgarella, V. Buat, J. Donas, B. Milliard, S. Chapelon}

\institute {Observatoire Astronomique Marseille-Provence, 
Laboratoire d'Astrophysique de Marseille, BP 8, 13376 Marseille 
Cedex 12, France }
\date{ Received 20 June 2000~/ accepted 11 January 2001 }
\offprints{D. Burgarella}
\mail{denis.burgarella@astrsp-mrs.fr}
\titlerunning{Visibility and quantitative 
morphology of galactic disks in rest-frame UV}
\authorrunning{Burgarella et al. }
\maketitle 
\begin{abstract} 

We have used ultraviolet (200~nm) images of the local spiral galaxies M33, M51,
M81, M100, M101 to compute morphological parameters of galactic disks at this
wavelength~:  half-light radius $r_{hl}$, surface brightness distributions,
asymmetries ($A$) and concentrations ($C_A$).  The visibility and the evolution of the
morphological parameters are studied as a function of the redshift.  The main
results are~:  local spiral galaxies would be hardly 
observed and classified if
projected at high redshifts (z $\ge$ 1) unless a strong luminosity evolution is
assumed.  Consequently, the non-detection of large galactic disks cannot be used
without caution as a constraint on the evolution of galatic disks.  Spiral
galaxies observed in ultraviolet appear more irregular since the
contribution from the young stellar population becomes predominent.  When these
galaxies are put in a (log~$A$ vs.  log~$C_A$) diagram, they move to the irregular
sector defined at visible wavelengths.  Moreover, the log~$A$ parameter 
is degenerate and cannot be used for an
efficient classification of morphological ultraviolet types. The analysis of 
high redshift galaxies cannot be carried out in a reliable way so far and a 
multi-wavelength approach is required if one does not want to misinterpret
the data.

\keywords {galaxies: evolution -- galaxies: fundamental parameters -- 
ultraviolet: galaxies -- galaxies: spiral} 
\end{abstract}


\section{Introduction}

At large distances, very concentrated galaxies with large surface brightnesses
are more easily observed. It is generally accepted that these galaxies are
probably spheroids (Steidel et al.  \cite{Steidel96}, Giavalisco et al.
\cite{Giavalisco96b}).  The detection of spirals and, more generally, galactic
disks which exhibit a lower surface brightness is far more difficult.  Moreover,
at high z we observe the rest frame ultraviolet (UV) emission of galaxies
redshifted in the visible.  If we wish to compare high redshift objects with
local ones, we must account for this effect by choosing templates observed in the
appropriate wavelength range.  The morphology of moderate and high z galaxies
has been intensively studied lately (Abraham et al. \cite{Abraham96a}; 
Abraham et al.
\cite{Abraham96b}; Schade et al.  \cite{Schade96}; Lilly et al.  \cite{Lilly98})
and structural parameters have been proposed which can be measured in a
rather automatic way (Abraham et al.  \cite{Abraham96a}).  These parameters are
quantities like the half-light radius ($r_{hl}$) or the mean surface brightness
but also more sophisticated quantities such as the concentration (C$_A$) or the 
asymmetry ($A$).
Before being applied to distant galaxies these parameters must be calibrated on
well-known templates which must be, as much as possible, representative of all the
galaxies expected at high z.  The calibration is generally made with catalogs of
nearby galaxies observed in the visible.  The sample of Frei et al.  (\cite{Frei96}) is
used largely for this aim (e.g.  in Abraham et al.  \cite{Abraham96a}; Conselice
et al.  \cite{Conselice00}; Bershady et al.  \cite{Bershady00}).

In trying to find an adequate tool to classify high-redshift galaxies, Abraham
et al.  (\cite{Abraham96b}) present a distribution of HDF galaxies in the log~A
versus log~$C_A$ plane. Abraham
et al.  (\cite{Abraham96b}) divide their diagram in three sectors calibrated at
z $\approx$ 0 and assimilated to E/SO, spirals and irreguliar/peculiar/merger galaxies
in agreement with van den Bergh et al. (\cite{vandenBergh96}).
The most important result is that the relative proportion of galaxies in the
three sectors seems to evolve~:  more galaxies lie in the
irr/pec/mrg area when moving to I$\ge$24 mag and the contribution from
large spiral galaxies is very close to zero at faint magnitudes. 
Although the redshift of
these galaxies is poorly constrained, Abraham et al.  (\cite{Abraham96b}) suggest
that the faint galaxies are mostly in the redshift range 0.5 $<$ z $<$ 2.5.
Combining morphological information with distance estimates, Driver et al.
(\cite {Driver98}) confirm the higher fraction of irregulars at low redshift
compared to the 1 $\le$ z $\le$ 3 range.  With the assumption that normal
galaxies are absent at high redshift, Driver et al.  (\cite {Driver98}) conclude
that the latter are the progenitors of the former.  However, Brinchmann et al.
(\cite{Brinchmann98}) performed simulations and observed an apparent migration
of galaxies towards later Hubble types which can be interpreted as a
misclassification of galaxies by about 24 \% at z $\approx$ 1.  Simulations were
also carried out by Abraham et al.  (\cite{Abraham96a}) by artificially
redshifting the Frei et al.  (\cite{Frei96}) sample of normal galaxies.  Only
a small number of these galaxies fall in the irr/pec/mrg area while most of
them lie in the spiral-E/SO (their dotted polygon).  Finally, Bunker et
al.  (\cite{Bunker00}) analyze the redshift evolution of high-redshift galaxies
directly from multi-wavelength data.  They compare the appearance of galaxies at
the same rest-frame wavelengths and find that morphological K-corrections are
generally not very important.  However, in the specific case of spiral galaxies,
the effect is more important and when the rest-frame wavelength moves to the UV,
the morphology does become more irregular.

As noted before (e.g.  Bohlin et al.  \cite{Bohlin91}; Kuchinski et al.
\cite{Kuchinski00}), it is necessary to take into account the apparent migration 
of spiral galaxies towards more irregular types in morphology-sensitive works. 
For instance,
works have been using the morphology classification of HDF galaxies to compute
morphology-dependent number-counts (Abraham et al.  \cite{Abraham96b}; Driver
et al.  \cite{Driver98}).  The misclassification of spiral galaxies due to
band-shifting is a strong bias that needs to be quantified before continuing in
the comparison of observations with models as underlined by Abraham et al.
(\cite{Abraham96a}).  The effect might be negligible at redshifts below z $\sim$
1 but as we will see, it becomes crucial when moving at redshifts of the order
of z $\ge$ 2. It will play a key role in the interpretation of 
future observations and in the
understanding of the formation and evolution of galaxies.

When redshifting nearby templates, Abraham et al. (\cite{Abraham96a}) have applied 
a {\sl K}-correction for each pixel
according to its color.  Here we adopt a more straightforward method by directly
redshifting UV images.  Pionneering work was carried out by Bohlin et al.
(\cite{Bohlin91}).  Kuchinsky et al.  (\cite{Kuchinski00}) has a similar approach by using
UIT Astro-2 images but no quantitative measurements have been performed on these
templates so far.

In this paper, we first study the morphology of some well-known local spiral
galaxies (M33, M51, M81, M100 and M101) to test their representativity.  Then,
we redshift these galaxies in the bands of the HST-WFPC2 (UBVRI) matching the
redshifts to remove any wavelength {\sl K}-correction.


\section{The galactic disk templates}

\begin{table*}
\caption[]{The templates galaxies. $\rm D_{25}$ are from Tully \& Richard (\cite{Tully88}), 
the distances of the galaxies~: D$_{M33}$ from Huterer et al. (\cite{Huterer95}), 
D$_{M51}$ from Feldmeier et al. (\cite{Feldmeier97}), D$_{M81}$ from 
Shara et al. 
(\cite{Shara99}), D$_{M100}$ from Ferrarese et al. (\cite{Ferrarese96}),
D$_{M101}$ from Stetson et al. (\cite{Stetson98}).
The ellipticity is measured with 
the software ELLIPSE of IRAF. The UV magnitudes have been measured on our images 
Note that M51 is a Seyfert~II galaxy and M81 an Active Galactic Nuclei galaxy. The morphological types are from Simbad~:
simbad.u-strasbg.fr/Simbad and the $B_T$ mag. from the LEDA database 
at www-obs.univ-lyon1.fr/leda/home\_leda.html.}
\begin{flushleft}
\begin{tabular}{ccccccccc} 

\hline 
Galaxy& Dist & $\rm M_B$& $\rm M_{UV}$& $\rm D_{25}$ & Ellipticity 
& morph. & pixel size \\

      & Mpc  & AB mag   & AB mag      & arcmin       &             
& type & arcsec.pixel$^{-1}$ \\ 
\hline
M 33 - NGC 598   & 0.88 & -18.6 & -15.7 & 56.5 & 0.4  &  Sc & 5.16 \\
M 51 - NGC 5194  & 8.4  & -21.0 & -18.4 & 13.6 & 0.25 &  Sc & 3.44 \\
M 81 - NGC 3031  & 3.5  & -20.1 & -16.0 & 22.1 & 0.45 &  Sb & 5.16 \\
M 100 - NGC 4321 & 16   & -21.2 & -18.5 & 6.1  & 0.05 &  Sc & 3.44 \\
M 101 - NGC 5457 & 7    & -21.2 & -19.2 & 23.8 & 0.10 &  Sc & 3.44 \\

\hline  
\end{tabular} 
\end{flushleft} 
\end{table*}

We have chosen to focus on five very well known galaxies.  They exhibit
different luminosities with -21 $\le$ $M_B$ $\le$ -18 and different star
formation activities.  All of them have been observed in UV by the FOCA
telescope.  The main characteristics of the galaxies are gathered in Tab.~1.

The FOCA telescope is a wide-field camera (Milliard et al.  \cite{Milliard92})
with a 150~nm-wide bandpass centered near 200~nm.  The camera is a
40cm Cassegrain telescope with an image intensifier coupled to a IIaO emulsion
film.  It was operated in two modes, the FOCA 1000 (f/2.6) and FOCA 1500
(f/3.8), which provide a 2.3 $\deg$-diameter field of view, 20
$\arcsec$-resolution, and 1.5 $\deg$-diameter field of view, 12
$\arcsec$-resolution, respectively.

The photometry was performed using the ELLIPSE software in IRAF.  The
ellipticities of the galaxies (Tab.~1) were estimated on
the images at z=0 and set fixed for the redshifted images for which only the
center was allowed to be adjusted. Given the low number of pixels in the  
redshifted images and the poor resolution on the disk we have not adjusted each 
isophote but prefered to adopt uniform values of P.A. and ellipticity. We have 
checked on the best detected cases that the results are not affected by this 
choice.


\section{Redshifting the galaxies}

\subsection{The method}
The way we processed our restframe-UV images to produce distant-like 
galaxies is similar to the method described in Giavalisco et al. 
(\cite{Giavalisco96a}). In brief, we rebinned the initial
image by a factor b defined as follows~:

$$b = {D (1+z)^2 \over L_z} {s_0 \over s_z} $$

\noindent where $L_z$ is the luminosity distance, D the real distance of the 
galaxy before
placing it to a redshift $z$ and $s_0$ and $s_z$ are the pixel sizes at 
$z \approx 0$ from the FOCA telescope (see Tab.~1) and at $z > 0$ 
from the HST WFPC2 camera (0.1 arcsec.pixel$^{-1}$) respectively. To compute the 
distance luminosity $L_z$, we used the redshift computed in Tab.~2 
($z = {\lambda_c / \lambda_z} - 1$), where $\lambda_c$ is the central 
wavelength of the
{\it HST} filter and $\lambda_z$ the wavelength of the emitted radiation. Here, 
$\lambda_z$=203~nm, which is the FOCA observation wavelength. In the following,
U will stand for the {\it HST} filter f336W, B for f439W, V for
f555W, R for f675W and I for f814W.

Note that we did not try to convolve
our images with the HST Point Spread Function (PSF). Indeed, even if the shape
of the PSF is well known (from short observations of stars close to the
center of the chips or from modelled PSFs), several effects are acting to prevent
us from obtaining a good accuracy. Observed PSFs vary with wavelength, time
and field positions. If we can deal with the first one, we have no specific reasons 
to select any values for the remaining ones (Holtzman et al. \cite{Holtzman95}). 
There is also evidence for sub-pixel Quantum Efficiency variation at the 10 \%
level. More realistic simulations to compare with specific observations might 
be obtained by using the appropriate PSF, but our goal is more generic. By not 
convolving our images, the effect is to produce images which are too ``peaky''. 
To give an order of magnitude, the light detected in the central pixel of a
non-resolved object would only be $\sim$ 70 \% of our value, the remaining would
spread over a 3x3-pixel area. Consequently, objects would be more difficult 
to detect in reality than in our simulations.

\begin{table}
\caption[]{The HST filters adopted when projecting local
galaxies at high redshifts. Col.~(1) is the name of the HST filter, col.~(2) 
is the central wavelength of the filter, col.~(3) the width of the band,  
col.~(4) the redshift corresponding to the band, assuming the restframe 
wavelength of FOCA at 203~nm and col.~(5) the calibration constant
computed with SYNPHOT in (erg.cm$^{-2}$.s$^{-1}$.${\rm 
\AA}^{-1}$)/(e$^{-}$.s$^{-1}$).}
\begin{flushleft}
\begin{tabular}{ccccc} 
\hline 
HST    &  $\lambda_c$  & $\Delta \lambda$ & z & calib. constant  \\
filter & (${\rm \AA}$) & (${\rm \AA}$)    &   &                  \\ 
\hline 
\hline 
 U-f336W & 3359.16 &  480.64 & 0.65 & 7.81 10$^{-18}$ \\
 B-f439W & 4311.84 &  476.37 & 1.12 & 4.12 10$^{-18}$ \\
 V-f555W & 5442.22 & 1229.96 & 1.68 & 4.90 10$^{-19}$ \\
 R-f675W & 6718.11 &  867.50 & 2.31 & 4.08 10$^{-19}$ \\
 I-f814W & 8001.60 & 1527.22 & 2.94 & 3.49 10$^{-19}$ \\
\hline  
\end{tabular} 
\end{flushleft} 
\end{table}

The average sky background estimated on the FOCA telescope is subtracted and, 
depending on the adopted redshifting scenario for evolution (next Sect.),
the galaxy is boosted or not. Next, the average pixel value $p_z$ at $z > 0$
is evaluated from the average pixel value $p_0$ at $z \approx 0$  
with the following formula~:

$$p_z = p_0 {a_0 \over a_z} {s_z^2 \over s_0^2} 
{\Delta \lambda_z \over \Delta \lambda_0} {1 \over {(1+z)^5}} {\rm t_{HST}}$$

where $a_0$ is the FOCA calibration constant and $a_z$ the {\it HST}
calibration constants given in Tab.~2. The dark current, estimated from the value given in
the WFPC2 Instrument
Hankbook v3.0 (0.005 e$^-$.s$^{-1}$.pixel$^{-1}$) and the sky background 
(23.3 V-mag.arcsec$^{-2}$ in agreement with sky values from the WFPC2 Instrument Handbook) 
are added to the image. The gain used throughout 
these simulations is 7 e$^-$/ADU. A poissonian noise and
a readout noise of 5 e$^-$.pixel$^{-1}$ are assumed. Note that the noise from the
original UV images is negligeable compared to the simulated noise (Figs. 1-5). 
36 images of each target were combined. The exposure time of individual images is 
1000 s with a total exposure time of $t_{HST}$ = 
10 hours. Figs.~1-5 present the results of the 
projection for our five galaxies. The luminosity evolution scenarios used 
are described in Sect.~3.2.

We can compare our HST limiting surface brightness with the data available in 
the literature and in the WFPC2 handbook. This is performed with the 
simulations using the f555W and f814W filters, since the variable uniform brightening
allows us
to scan the S/N scale (everything else kept constant~: exposure times,
etc.) . The 1-$\sigma$ limiting magnitude is $\mu_{AB}$ = 26.5 
mag.arcsec$^{-2}$ for our galaxies, for the adopted 10h-exposure time and 
with the chosen instrumental configuration. For such a surface 
brightness, and with the above filters and 45$^o$ declination (as assumed 
in our simulations), the WFPC-2 Exposure Time Calculator (ETC) returns a S/N 
$\approx$ 1.5 per pixel, in reasonable agreement with our computation. We can 
also compare these values with the 1$\sigma$ limiting isophote of the HDF data 
of 26.5 mag.arcsec$^{-2}$ quoted by Abraham et al. (\cite{Abraham96b}). Given a 
f606W HDF exposure time of $\sim$ 35h and assuming the S/N scales as the square 
root of the exposure time, we should have a limiting surface brightness of 
$\sim$ 25.8 mag.arcsec$^{-2}$, which is slightly brighter but still consistent 
with our values. On the other hand, Giavalisco et al. (\cite{Giavalisco96b}) 
presents a limiting surface brightness of 29.31 mag.arcsec$^{-2}$ in the f606W 
for an exposure time of 15600 sec., which is much dimmer. We have no clear 
explanation for this discrepancy.

\begin{figure}
\resizebox{8cm}{8cm}{\includegraphics{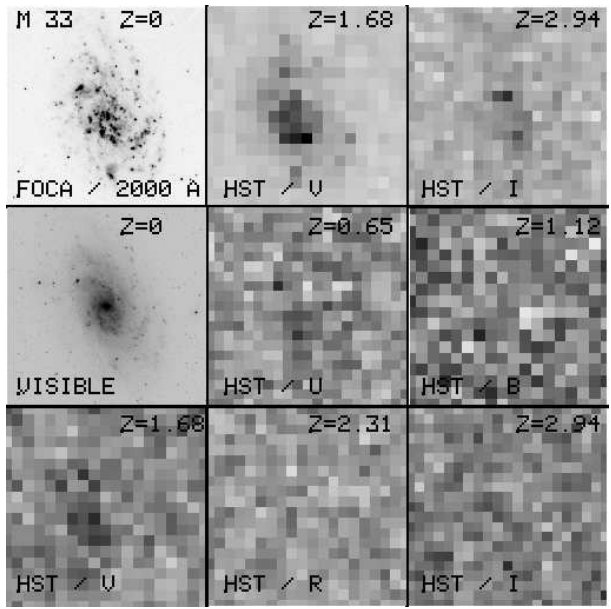}}
\caption[]{From top to bottom and left to right, this montage presents 
sequentially the ultraviolet image of M33 at z=0, the redshifted images in V 
and I uniformely boosted by 4 mag (scenario 3), the visible image of M33 
(from the DSS) and the redshifted (scenario 2) images in the UBVRI filters 
of the WFPC2 (see text and Tab.~2.). Note that these images are not convolved
with the {\it HST} PSF.}
\label{viewm33}\end{figure}

\begin{figure}
\resizebox{8cm}{8cm}{\includegraphics{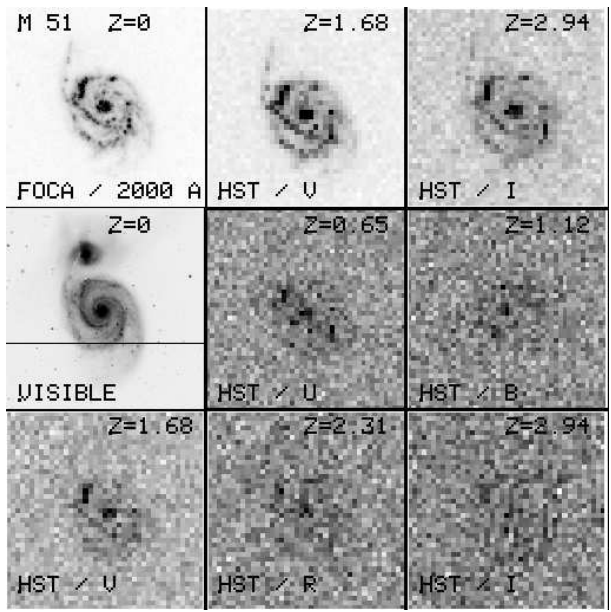}}
\caption[]{Idem for M51.}
\label{viewm51}\end{figure}

\begin{figure}
\resizebox{8cm}{8cm}{\includegraphics{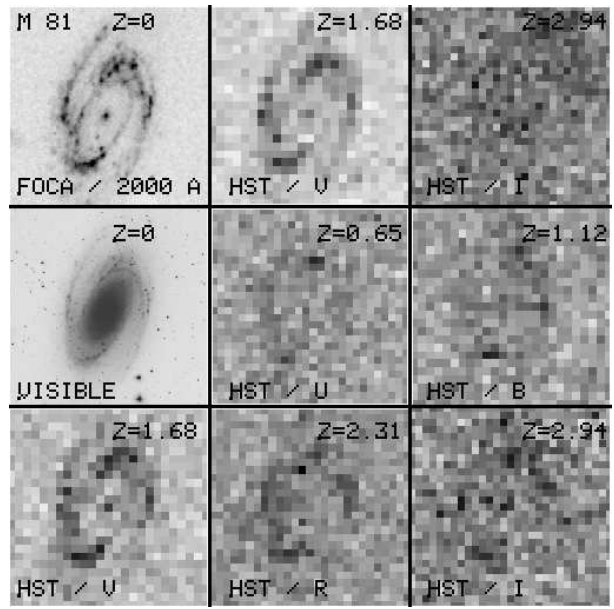}}
\caption[]{Idem for M81.}
\label{viewm81}\end{figure}

\begin{figure}
\resizebox{8cm}{8cm}{\includegraphics{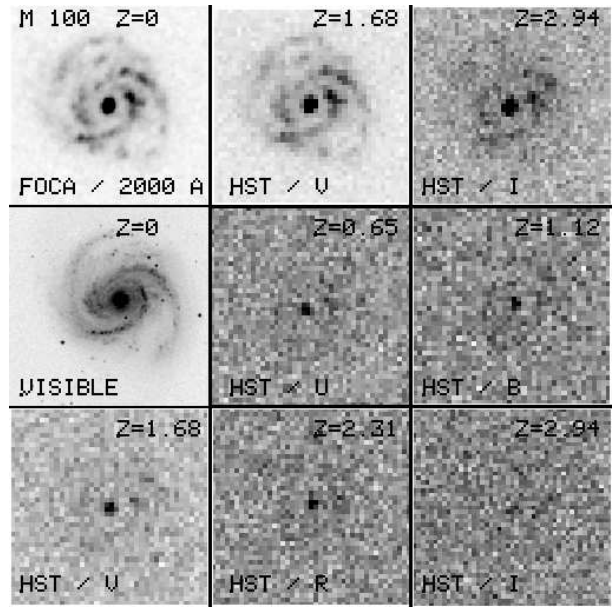}}
\caption[]{Idem for M100.}
\label{viewm100}\end{figure}

\begin{figure}
\resizebox{8cm}{8cm}{\includegraphics{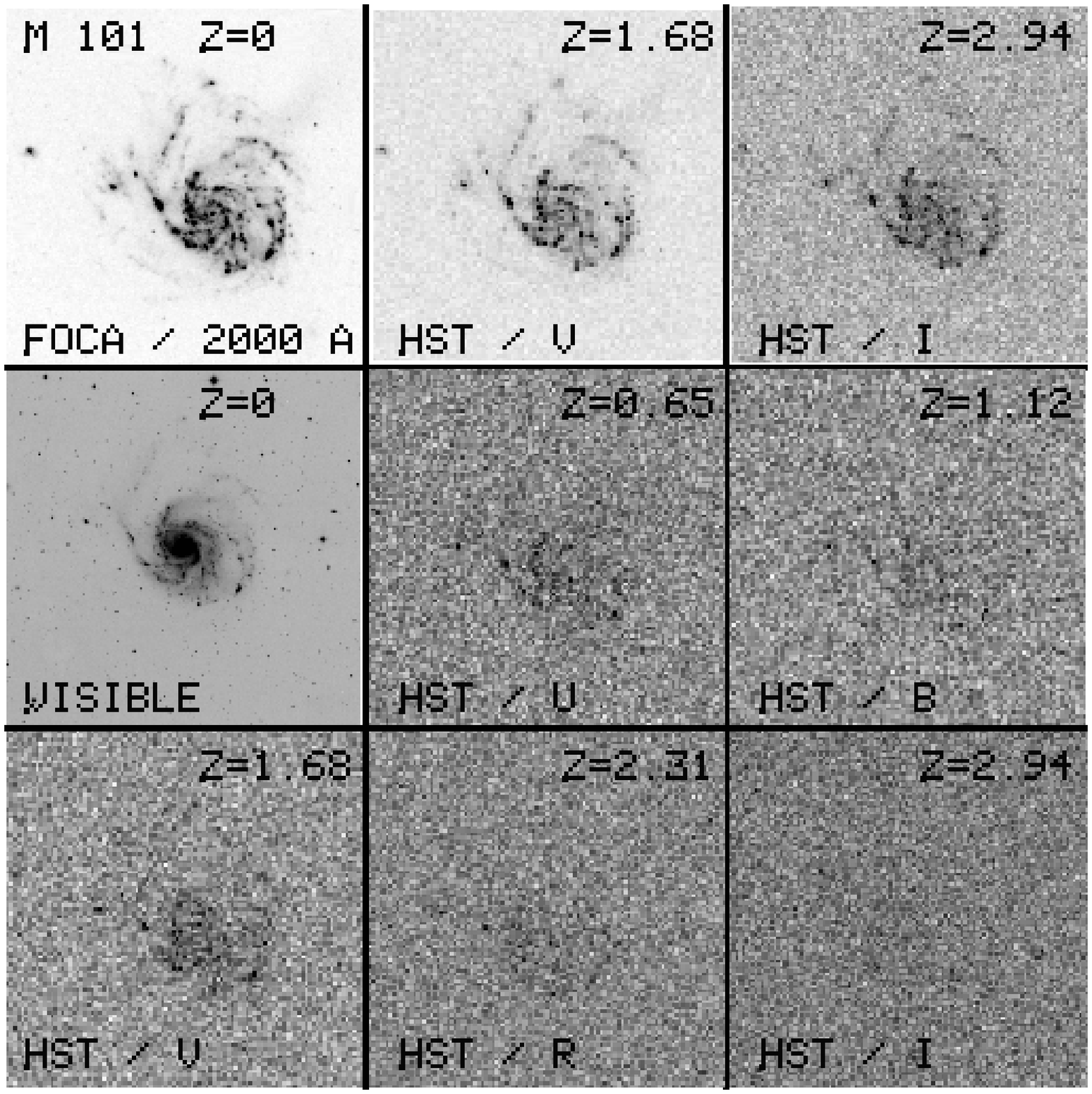}}
\caption[]{Idem for M101.}
\label{viewm101}\end{figure}

\subsection{The adopted evolutionary scenarios}

Three scenarios have been adopted to move the galaxies away and
simulate younger galaxies.  First, we
simply redshifted them without any modification :  no evolution (scenario~1).
As we will see below this scenario leads to almost no detections, even at
moderate redshifts.  Therefore, we assumed some evolution (scenario~2): 
we adopted
an exponential decrease of the star formation rate with an e-folding rate of 8
Gyrs except for M81 which has a e-folding rate of 3 Gyrs.  These
values are consistent with those expected from the morphological types of the
galaxies (e.g.  Kennicutt et al.  \cite{Kennicutt94}).  The adopted 
evolutionary scenarios translate into a higher UV magnitude when we simulate
younger galaxies.
The increases in magnitude due to evolution vary with the redshift 
by 2.5, 3.2, 3.6, 3.9 and 4 mag in 
U, B, V, R and I respectively for M81 and by 0.9, 1.2, 1.4, 1.5, 1.5 mag in 
U, B, V, R and
I respectively for the other galaxies.  However, this evolutionary scenario is
not very efficient for the detection of galaxies.  For 
the purpose of actually seeing galaxies and estimating the magnitude needed
to observe them, we also applied arbitrary
luminosity increases to each pixel of the galaxies from 1 to 4 mag to all the
 galaxies (see Sect.~5). This last scenario (scenario~3) was
only applied to the V-band and I-band or equivalently to the galaxies 
redshifted to z=1.68  and  z=2.94.


\section{ Detailed properties of the galactic disks}
\subsection{The surface brightness}

\begin{figure} \resizebox{\hsize}{8cm}{\includegraphics{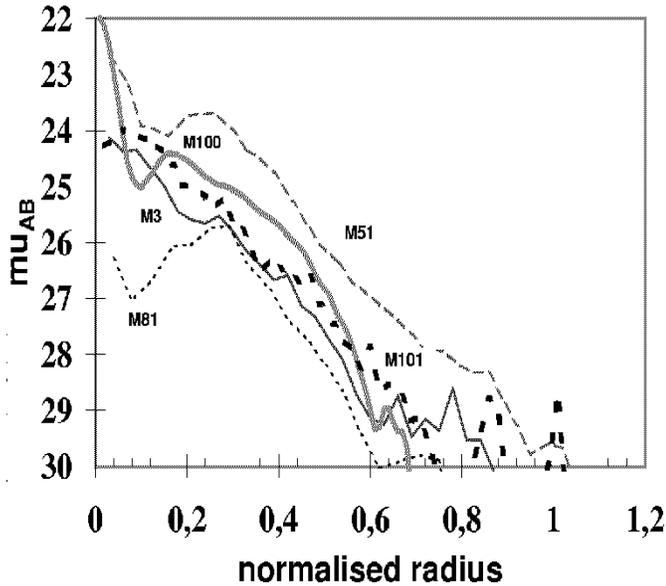}}
\caption[]{The surface brightness profiles of our galaxy
sample observed in UV at z $\approx$ 0. Note that the radii are normalized 
(see text)} \label{RmuAB} \end{figure}

We measured the surface brightness distribution of each galaxy in UV (at 
z $\approx$ 0) and at different redshifts. The UV rest-frame surface brightnesses
are presented in Fig.~\ref{RmuAB}.
For comparison, we normalized them to the semi-major axis 
of the total aperture used for the photometry and defined in
section 4.2.1.  Whereas M33 and M101 exhibit a rather linear profile, as expected
for exponential disks, M51 and M81 have a non-monotonic distribution and M100
may be viewed as an intermediate case.  Such a
difference in radial profiles will lead to variations in the morphological
classification as discussed below.  No clear bulge is present in the profiles.  
Since no recent star formation appears in the bulge, it disappears in UV. These 
typical morphological changes have been already described in Kuchinski et al. 
(\cite{Kuchinski00}).
M51 is classified as a Seyfert~2 galaxy and M81 as an AGN galaxy 
but the contribution from the nucleus to 
the overall UV emission is not important. On the other hand, there is some UV 
light in
the core of M100 which seems to be produced within a nuclear star-formation
region (e.g Ryder \& Knappen \cite{Ryder99}).

We also calculated the UV surface brightness within the ellipse which
encloses half the total light.  The measured values are in very good
agreement with the predicted values (e.g. Lilly et al.'s  \cite{Lilly98})~:

$$\rm \mu_{AB} (z) = \mu_{AB} (RF)+2.5 \log(1+z)^3-BST$$

\noindent $\rm \mu_{AB} (z)$ is the surface brightness observed in the HST filters, $\rm
\mu_{AB} (RF)$ the rest frame surface brightness measured in the UV image and
BST the boost in magnitude which varies according to the adopted scenario.  In
this formula we only account for the cosmological dimming term without $\sl
K$-corrections since the redshifts were chosen to match the {\it HST} filters and
to avoid these $\sl K$-corrections.

\subsection{Morphological parameters}

\subsubsection{The half-light radius}

The half-light radius ($r_{hl}$) is a basic parameter which measures the size of
a galaxy.  The critical point is to define the total aperture of the galaxies.
We adopt the method detailed in Bershady et al.  (\cite{Bershady00}) which 
defines the total
aperture to perform photometry as twice the semi-major axis (called hereafter
the major radius) where $\rm \eta(r) = I(r)/<I(r)> = 0.2$.  Note that $I(r)$ is
the local surface brightness and $\rm <I(r)>$ the average surface brightness
within the major radius $r$.  With such a definition we avoid the need to define
isophotal radii which are redshift dependent.  In practice, the total flux thus
estimated is similar to that deduced from the analysis of the curve of growth.
The total UV magnitude of the galaxies at z $\approx$ 0 are reported in Tab.~1.

The half-light major radius was measured in each detected galaxy and the
results are reported in Tabs.~3-7.  Predicted values were calculated with the
measurement at z$\approx$0 and are also reported in these Tables.  A remarquable
agreement is found between the measured and predicted quantities.  In physical
units this corresponds to 6 kpc for M81 and M51, 6.8 kpc for M100, 13 kpc for
M101 and 3.8 kpc for M33.  M101 appears to be very extended:  its half-light
radius is approximately twice that of M51.  As a comparison the ratio of their
diameter at B=25 mag ($D_{25}$) is only 1.45.  In UV M101 is 0.8 mag brighter
than M51 while the difference is of 0.5 mag in B (Tab.~1).  M101 is a diffuse
object whereas M51 has a very high surface brightness.  As expected the
detection of M51 at high z is far easier than that of M101 in spite of their
absolute luminosity.  M81 and M33 are intrinsically fainter with $M_{UV}$ $>$
-18.5 but here again their half-light radii are very different due to their very
different UV distribution.

Our conclusion is that the half-light radius appears as a very robust parameter
when the galaxy is moved away at high z but only gives a rough
estimate of the galaxy size without any morphological information.

\subsubsection{The concentration}

{\it Kent's parameter}

The concentration parameter is a classical quantity first introduced by
Kent (\cite{Kent85}). It is generally defined as the logarithm of the ratio of two
radii~:  

$$\rm C_K = 5 \log(r_{0.8}/r_{0.2})$$ 

where $r_{0.8}$ is the outer radius enclosing 80$\%$ of the total flux and 
$r_{0.2}$ is the inner radius enclosing 20$\%$ of the total flux.  Such a
definition is not based on isophotes and is therefore not dependent on surface
brightness dimming as soon as the total aperture to perform photometry is
defined independently of isophotal levels.  Bershady et al. (\cite{Bershady00}) 
have found
that this parameter is remarkably stable against resolution degradation and
conclude that it is very suitable for high redshift measurements.\\

We calculated the concentration parameter for all the galaxies detected.
The aperture to determine the total flux was defined as in the previous
section.  As expected it
appears very stable when the galaxies are redshifted, varying by less than 0.1 as
soon as the signal-to-noise ratio (defined within the half-light radius) is
larger than 20.  However, the absolute values of $C_K$ are out of the range
usually found for disk galaxies.  All the objects exhibit a very low value of
$C_K$ from 1.5 to 2.6 with $C_K(M51)= 2$, $C_K(M81)=1.5$, $C_K(M101)=2.6$, 
$C_K(M100)=2.3$ and $C_K(M33)=2.5$ whereas typical
values for galactic disks are larger than 3, even for late-type disk galaxies
(Kent \cite{Kent85}, Bershady et al. \cite{Bershady00}).  \\

This result must be related to the surface brightness profiles presented in
Fig.~\ref{RmuAB}.  The value of $C_K$ for an exponential disk is 2.7, in agreement
with the values found for M33 and M101 whose distributions look like
exponentials.  The very low values found for M51 and M81 are due to their
irregular UV distribution with a low central emission and a bright annulus .
As already underlined, no bulge is visible. For the five galaxies $C_K$ is
found too low, due to the absence of a bulge in UV.  This result lowers the
importance of this parameter for high redshift galaxies observed in a UV rest
frame unless a reliable calibration on a large database of templates of all
types of galaxies is performed.  The calibration made with the Frei sample in B
and R are not representative of the UV morphology.

{\it Abraham et al.'s parameter}

More recently, Abraham et al.  (\cite{Abraham96a} and references therein) have
introduced another definition of the concentration as the ratio of fluxes within
two isophotal radii.  The outer galaxy isophote is fixed at a given level above
the sky (1.5 or 2 $\sigma$) and the inner isophote is defined as having a radius
equal to 0.3 times the radius of the outer isophote.  The concentration
parameter is the ratio of the fluxes between these inner and outer isophotes.
Here we adopt the definition:  $$C_A = {\sum_{E_\alpha}
I_{ij}\over{\sum_{E_{iso}} I_{ij}}}$$ where $\rm E_{iso}$ refers to the
elliptical aperture with a semi major axis ($sma_{iso}$) corresponding to the
outer isophote at 1.5 or 2 $\sigma$ and $E_{\alpha}$ to the elliptical aperture
with the semi major axis equal to $0.3 \cdot sma_{iso}$ (see Fig.~\ref{RmuAB}).

This parameter depends on isophotes and is therefore subject to potential
problems since the surface brightness is a steep function of the redshift.  This
difficulty has lead Bershady et al.  to prefer the concentration defined by
Kent.  Brinchmann et al.  (\cite{Brinchmann98}) chose to use Abraham et al.'s
concentration (hereafter $C_A$) but with a correction of this effect.  They
assume a de Vaucouleurs law to perform their correction.  Such a distribution is
certainly not valid for the UV distribution of star forming galaxies.  Here we
adopt a more empirical approach by directly measuring the concentration on
redshifted images of real nearby galaxies.

$C_A$ was measured in UV at z $\approx$ 0 for an isophotal level of 1.5 and
reported in Tabs.~3-7.  The concentration is calculated only when the isophote
at the adopted level is closed.  All the values found are lower than 0.4
($\log(C_A)\le -0.4$) which is characteristic of spirals and irregulars (e.g.
Abraham et al.  \cite{Abraham96a}).  M81 appears very extreme:  this quiescent
early-type spiral has a very low concentration.  Moreover, the concentration
parameter $C_A$ of each galaxy is stable when the galaxy is redshifted.  The
difference $\Delta C_A = |C_A(z=0) - C_A(z \approx 0) |$ $\le$ 0.16 for all the
spiral galaxies studied here except for one value for M33 (scenario 3 and boost
by 4 mag) which has a large uncertainty.

Therefore Abraham et al.'s concentration parameter appears as a robust one.  It
is more adequate than Kent's one to describe the UV morphologies at low and high
redshifts.

\begin{table}
\caption[]{Detectability and morphology of Messier 51. In the first part of the 
table, the galaxy is redshifted assuming an exponential evolution with $\tau =$ 
8 Gyr, in the second and third 
parts of the table the galaxy is boosted by a constant factor (1 to 4 mag) and 
redshifted in the V band (z=1.68) and in the I band (z=2.94). The concentration 
and asymmetry are defined as by
Abraham et al. The concentration is calculated at the 1.5$\sigma$ level. The symbol
``:'' after a value points out its large uncertainty, $SB$ means surface brightness
within the half-light radius $r_{hl}$ in m$_{AB}$/arcsec$^2$ and $\Delta$ V and 
$\Delta$ I the brightening in V and I. The symbol ``nd'' means that the galaxy
was not detected in this filter}
\begin{flushleft}
\begin{tabular}{lllllll} 

\hline 
Filter  & UV & U    & B    &V     & R    & I \\
\hline 
z          & 0  & 0.65 & 1.12 & 1.68 & 2.31 & 2.94 \\
S/N        &$5000$& 18 &13& 26& 15& 9\\
$\rm m_{AB}$& 11.2&23.5&24.0&24.9&25.1&25.3\\
$\rm r_{hl}(")$&151&0.75&0.75&0.70&0.78&0.84\\
pred. $\rm r_{hl}$ & &0.77 &0.73 &0.72 &0.77 &0.83\\
$A$        & 0.26&0.3:&0.4:&0.31&0.35&0.4:\\
$C_A$      &0.29&0.17&  &0.17&    &   \\
SB         &23.7&24.5&25&25.7&26.2&26.5\\
\hline 
\hline 
$\Delta$ V & 0 mag & 1 mag & 2 mag & 3 mag & 4 mag \\
\hline 
S/N &8 &20 &50 &130 & 300\\
$\rm m_{AB}$&26&25&24&23&22\\
$\rm r_{hl}(")$&0.61:&0.79&0.73&0.75&0.73\\
$A$&  0.5:& 0.27&0.30&0.28&0.26\\
$C_A$& &0.17&0.18&0.23&0.25\\
SB &26.8&26&25&24&23\\
\hline  
\hline 
$\Delta$ I & 0 mag & 1 mag & 2 mag & 3 mag & 4 mag \\
\hline 
S/N & nd&7 &12&28 & 70\\
$\rm m_{AB}$& &25.6&24.9&24&23\\
$\rm r_{hl}(")$& &0.73:&0.8&0.8&0.8\\
$A$&  & 0.6:&0.35&0.28&0.25\\
$C_A$& &&& 0.17&0.17\\
SB & &26.5&26&25.2&24.2\\
\hline  
\end{tabular} 
\end{flushleft} 
\end{table}

\begin{table}
\caption[]{Detectability and morphology of Messier 81. Same 
as table 3 but with $\tau$ = 3 Gyrs}
\begin{flushleft}
\begin{tabular}{lllllll} 
\hline 
Filter& UV & U & B &V & R & I \\
\hline 
z   & 0 & 0.65 & 1.12 & 1.68 & 2.31 & 2.94 \\
S/N &$3000$& 10 & 10& 30& 20& 10\\
$\rm m_{AB}$& 11.7&24.3&24.3&24.5&24.6&25.2\\
$\rm r_{hl}(")$&366&0.76&0.72&0.79&0.87&0.8\\
pred. $\rm r_{hl}$ & &0.79 &0.72 &0.74 &0.78 &0.84\\
$A$& 0.21& 0.4:&0.5: &0.32&0.36 & 0.3:\\
$C_A$& 0.09 & & &0.08& 0.05 & \\
SB &26&25.3&25.2&25.6&25.8&26.3\\
\hline  
\hline 
$\Delta$ V & 0 mag & 1 mag & 2 mag & 3 mag & 4 mag \\
\hline 
S/N &nd& nd& 8&20  & 45\\
$\rm m_{AB}$&&&26.5:&25.3&24.4\\
$\rm r_{hl}(")$& & & 0.7:& 0.75&0.73 \\
$A$&  & & 0.4:&0.38 & 0.25\\
$C_A$& & & &0.06& 0.09\\
SB & & &27&26.2&25.2\\
\hline  
\hline 
$\Delta$ I & 0 mag & 1 mag & 2 mag & 3 mag & 4 mag \\
\hline 
S/N &nd& nd& nd& nd& 8\\
$\rm m_{AB}$&&&&&25.3\\
$\rm r_{hl}(")$& & &&&0.8 \\
$A$&  & & & &0.3: \\
$C_A$& & & & & \\
SB & & & & &26.4\\
\hline  
\end{tabular} 
\end{flushleft} 
\end{table}

\begin{table}
\caption[]{Detectability and morphology of Messier 101. Same as table 3. }
\begin{flushleft}
\begin{tabular}{lllllll} 

\hline 
Filter  & UV & U & B &V & R & I \\
\hline 
z   & 0 & 0.65 & 1.12 & 1.68 & 2.31 & 2.94 \\
S/N &$5000$& 20& 14&26 & 15& nd \\
$\rm m_{AB}$&10&22.5:&23.4:&24&24.4:&\\
$\rm r_{hl}(")$&384&1.63:& 1.4:& 1.56&1.4: &\\
pred. $\rm r_{hl}$  & &1.66&1.52&1.54&1.65 &1.78\\
$A$& 0.40&0.39 &0.5: &0.41 & &\\
$C_A$&0.30& & &0.19 & & \\
SB &24.5&25.4&25.9&26.5&27& \\
\hline 
\hline 
$\Delta$ V & 0 mag & 1 mag & 2 mag & 3 mag & 4 mag \\
\hline 
S/N &10 &20 &47 &115 & 290\\
$\rm m_{AB}$&25.7: &24.3&23.3&22.2&21.1\\
$\rm r_{hl}(")$&1.1: & 1.3:&1.5&1.45&1.5 \\
$A$&  &0.6: &0.33&0.40&0.40\\
$C_A$& & 0.17&0.14& 0.25& 0.30\\
SB & 27.8: &26.8&25.9&24.9&23.9\\
\hline 
\hline 
$\Delta$ I & 0 mag & 1 mag & 2 mag & 3 mag & 4 mag \\
\hline 
S/N &nd& nd& 10&25  & 64\\
$\rm m_{AB}$& & & 24.1&23.1&22\\
$\rm r_{hl}(")$& & & 1.7:& 1.86&1.9 \\
$A$&  & & 0.5:& 0.42& 0.39\\
$C_A$& & & & & 0.14\\
SB & & &27.3 &26.3 &25.3 \\
\hline
\end{tabular} 
\end{flushleft} 
\end{table}

\begin{table}
\caption[]{Detectability and morphology of Messier 33. Same as table 3.}
\begin{flushleft}
\begin{tabular}{lllllll} 

\hline 
Filter  & UV & U & B &V & R & I \\
\hline 
z   & 0 & 0.65 & 1.12 & 1.68 & 2.31 & 2.94 \\
S/N &$5500$& 3& nd & 6& nd&nd\\
$\rm m_{AB}$&9.0&26.2& &26.9& \\
$\rm r_{hl}(")$&889&0.4: & &0.4: & &\\
pred. $\rm r_{hl}$  & &0.49&0.44&0.45& 0.48&0.52\\
$A$& 0.31& & &0.6: & &\\
$C_A$&0.33& & & & & \\
SB s&25&26& &26.7& &\\
\hline 
\hline 
$\Delta$ V  & 0 mag & 1 mag & 2 mag & 3 mag & 4 mag \\
\hline 
S/N &nd & 4&10 & 26&60 \\
$\rm m_{AB}$& &27.4&26.5&25.5&24.5\\
$\rm r_{hl}(")$& &0.3: &0.4:&0.38&0.40 \\
$A$&  & &0.5: &0.5: &0.5:\\
$C_A$& & & &0.23&0.24\\
SB & &26.7&25.9 &24.9& 23.9\\
\hline
\hline   
$\Delta$ I  & 0 mag & 1 mag & 2 mag & 3 mag & 4 mag \\
\hline 
S/N &nd & nd& nd&5 &13\\
$\rm m_{AB}$&&&&26.6&25.5\\
$\rm r_{hl}(")$& & &&0.44&0.55 \\
$A$&  & & & 0.6:&0.4:\\
$C_A$& & & &&0.06:\\
SB & && &26.3&25.5 \\
\hline
\end{tabular} 
\end{flushleft} 
\end{table}

\subsubsection{The asymmetry}

Asymmetry is one of the most natural way of analyzing morphology and 
classifying galaxies. Conselice et al. (\cite{Conselice00}) present a detailed 
study of rotational asymmetry in galaxies. Here, we define the asymmetry in the
same way as Abraham et al. (\cite{Abraham96a})~:

$$ A = 1/2 (min[{\Sigma|I_0 - I_{\Phi}| \over \Sigma |I_0|}] - 
       min[{\Sigma|B_0 - B_{\Phi}| \over \Sigma |I_0|}]) $$
       
\noindent where $I$ represents the image pixel values and $B$ the background 
pixel values. The rotation angle $\Phi$ is set to 180 $\deg$ in this paper, 
which means that the I$_{\Phi}$ images are rotated by 180 $\deg$ before  
subtraction with the original image. The first term on the right side of the 
equation corresponds to the asymmetry of the source. Note, 
however, that the rotational asymmetry is a measurement based on individual 
pixels and noise poses an important problem. Consequently, Conselice
et al. (\cite{Conselice00}) introduce a noise correction (only valid for 
uncorrelated noise and therefore not for HDF dithered images) which is computed in the 
latter term. This correction consists of estimating the asymmetry of blank areas
in the neighborhood of the galaxy. In order to optimize this calculus we must 
check that the computed 
asymmetry is really at a minimum and an additional step is to compute $A$ 
at different positions on a grid and keep the minimum value. Another key-point 
lies in the signal-to-noise ratio. In this paper, we computed $A$ for 
all detected galaxies. In agreement with Conselice et al. 
(\cite{Conselice00}), we computed the asymmetry up to the radius where 
$\eta$ = 0.2 ($\eta(r)$ = ${I(r) \over <I(r)>}$), which permits us to define a 
maximum radius independent of the distance/redshift and of the photometric 
calibration.
The half-light integrated S/N must exceed $\sim$ 20 in order to have reliable 
estimates for A. This is less than the limiting S/N values
reached by Conselice et al. (\cite{Conselice00}). As expected, the galaxies 
appear
very asymmetric with $A>0.2$. This point will be discussed in Sect.~6.

\begin{table}
\caption[]{Detectability and morphology of Messier 100. Same as table 3. }
\begin{flushleft}
\begin{tabular}{lllllll} 

\hline 
Filter  & UV & U & B &V & R & I \\
\hline 
z   & 0 & 0.65 & 1.12 & 1.68 & 2.31 & 2.94 \\
S/N         &1600& 14 &   10 & 20 &11& nd\\
$\rm m_{AB}$&12.6& 23.6 & 24.2 & 24.6 &25.2& \\
$\rm r_{hl}(")$&88&0.82 &0.82 &0.88 &0.8: &\\
pred. $\rm r_{hl}$ & &0.90&0.79&0.81&0.86 &0.93\\
$A$& 0.20& 0.45& 0.5:&0.37 & 0.45&\\
$C_A$&0.25& & & & & \\
SB &24.4&25.1& 25.6&26.3& 26.7&\\
\hline 
\hline 
$\Delta$ V  & 0 mag & 1 mag & 2 mag & 3 mag & 4 mag \\
\hline 
S/N &nd & 15&35 & 83&210 \\
$\rm m_{AB}$& &27.4&26.5&25.5&24.5\\
$\rm r_{hl}(")$& &0.9: &0.82&0.82&0.84 \\
$A$&  & 0.4:& 0.24& 0.22&0.22\\
$C_A$& & &0.21 &0.21&0.22\\
SB & &26.8:&25.7 &24.7& 23.7\\
\hline
\hline   
$\Delta$ I  & 0 mag & 1 mag & 2 mag & 3 mag & 4 mag \\
\hline 
S/N &nd & nd& 7&20 &50\\
$\rm m_{AB}$&&&25.2:&24.1&23.1\\
$\rm r_{hl}(")$& & &0.9&0.98& 1.0\\
$A$&  & & &0.30 &0.20\\
$C_A$& & & &&0.21\\
SB & && 26.9&25.9& 25\\
\hline
\end{tabular} 
\end{flushleft} 
\end{table}


\section{Detection of disks at high redshifts}

The first question that we will address is the
detectability of disks at high redshifts. 
The evolution of the B mean surface brightness of large disk-dominated galaxies
was thought to increase by a value of $\Delta \mu$ ranging from 0.8 to 1.6 mag
between now and a redshift of z $\approx$ 1 compared to Freeman's
(\cite{Freeman70}) value at z = 0 (Schade et al.  \cite{Schade96}; Lilly et al.
\cite{Lilly98}; Roche et al.  \cite{Roche98}, Bouwens \& Silk \cite{Bouwens00}).
However, Simard et al.  (\cite{Simard99}) performed a similar analysis but took into
account a selection function in the magnitude-size plane as a function of
redshift.  The main effect produced by the above bias is that galaxies with low
surface brightnesses are lost at high redshit.  Before correction, the mean disk
surface brightness would increase by $\Delta \mu$ = 1.3 mag from z=0.1 to z=0.9
consistently with the values estimated by Schade et al.  (\cite{Schade96}),
Lilly et al.  (\cite{Lilly98}), Roche et al.  (\cite{Roche98}).  After
accounting for the selection effect, no discernible evolution is observed in the
disk surface brightness of disk-dominated galaxies brighter than $M_B$ = -19.
However, using the same dataset Bouwens \& Silk (\cite{Bouwens00}) reach a
different conclusion ($\Delta \mu$ $\sim$1.5 mag of evolution).  This difference
stems from the fact that Bouwens \& Silk (\cite{Bouwens00}) find, in the
observation, a number of high surface brightness galaxies exceeding model
predictions.  These authors therefore argue that there is a strong evolution in
the total number of high surface brightness galaxies from z = 0 to z $\approx$ 1
not accounted for by Simard et al.  (\cite{Simard99}).

We can analyze the detectability of our simulated high-redshift observations in
HST bands.  The first point to note is that M33, M81 and M100 become
undetectable if they are redshifted in the {\it HST} V-band (z=1.68) without any
other modifications.  The situation is only more favourable for M51 and M101
which are detected in the V-band with an integrated S/N
$\approx$ 10.  Note that the S/N are measured within ellipses which enclose half
the total light of the galaxy.  Even when detected, these low S/N prevent any
safe estimations of morphological parameters as discussed below.  None of the
galaxies are detected in the {\it HST} I-band (z=2.94).  The results for the
V-band and I-band without evolution are reported in Tabs.~3-7 (referred to as a
brightening of 0 mag).  The V-band appears as the best configuration to maximize
S/N as it combines a rather moderate redshift (z=1.68) with an efficient filter
(f555W). 

In the following we will use scenario~2 of luminosity evolution
presented in Sect.~3.2 for the galaxies and analyze their effect on the
detectability of the spiral galaxies. For all the galaxies other than M81, this scenario 
implies an evolution of the surface brightness consistent with the values
found in the literature and presented at the beginning of the section. 
The S/N barely reaches 30, i.e.  below
the value of 50 which is considered by Conselice et al.  (\cite{Conselice00}) as
necessary to begin to estimate reliable morphological parameters.  It seems,
however that some reasonably safe work can be carried out down to S/N $\approx$
20-30, depending on the galaxies (see Sect.~5.2).  The adopted evolutionary
scenario is not enough for M33 and S/N $<$ 10.  For the other spiral
galaxies, 10 $\le$ S/N $\le$ 30.  Nevertheless, in only a limited number of
cases, $A$ and $C_A$ can be safely estimated.  If such galaxies are observed at
high redshift, it would be often impossible to perform a reliable morphological
analysis.

In order to constraint the evolution needed to detect our local
spiral galaxies and estimate their morphological parameters, we applied
scenario~3, i.e.  a uniform magnification (same one to each pixel of the image)
ranging from 1 to 4 magnitudes to our
galaxies observed at z=1.68 and z=2.94 i.e.  observations through the WFPC2
V-f555W and I-f814W filters.  In the V-band, S/N $\ge$ 20 is the minimum needed
to perform any morphological analysis.  M51 reaches such a value with a
boost of 1~mag. This boost is slightly less that the V-boost adopted in scenario~2
and explains the positive results for this filter. Assuming H$_0$=50 
km.s$^{-1}$.Mpc$^{-1}$ and q$_0$=0.5, this translates into an e-folding rate 
slightly larger than in scenario~2~: $\tau_V=10.9$ Gyrs.

The high surface brightness of M51 is the major characteristic
that helps detection of this spiral galaxy. A boost by 2~mag (i.e.  $\tau_V=5.5$ Gyrs)
is needed for M100 and M101 to get reliable morphological parameters. 
Up to 3~mag (i.e.  $\tau_V=3.6$ Gyrs) and more than 4~mag (i.e. $\tau_V=2.7$ 
Gyrs) are necessary to measure $A$ and $C_A$ 
in V.  The situation is slightly worse in the I-band
where a reliable estimate of the morphology corresponds to a boost by 3~mag 
($\tau_I=4.1$ Gyrs) for M51, by 4~mag ($\tau_I=3.1$ Gyrs) for M100
and M101 with the same assumptions on the
cosmology.  M33 and M81 are never detected in I, which implies boost $>$ 4~mag
(i.e. $\tau_I~>~3$ Gyrs) for a detection. Such evolutions are very high and
imply e-folding rates more typical of very early type galaxies dominated or
largely influenced by the bulge component. 
Roche et al. (\cite{Roche98}) found $\sim$ 2.8~mag of surface brightness 
evolution for galaxies at 2 $<$ z $<$ 3.5 relative to galaxies at z $<$ 0.35.

\begin{figure}
\resizebox{8cm}{8cm}{\includegraphics{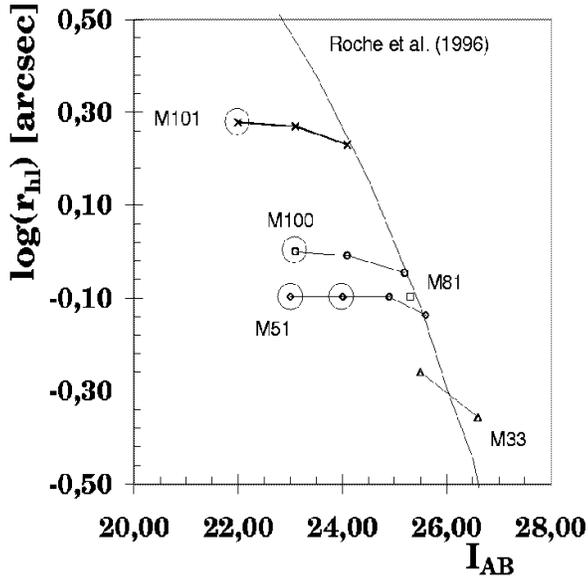}}
\caption[]{The location of the galaxies when detected
in the I-band in the log of the half-light radius log($r_{hl}$) in arcsecs vs. the
$I_{AB}$ magnitude within $r_{hl}$. The line corresponds to the detection limit
modelled by Roche et al. (\cite{Roche96}) and scaled to $I_{AB}$. 
 It can be seen than M33 and M81 are only marginally detectable within
our modelled evolution. M51, M100 and M101 can be detected with lower boosts
but their morphology cannot be measured except for the major boosts which are
shown with large circles around the symbols.}
\label{logrhlIab} \end{figure}

In conclusion, except perhaps for spiral galaxies with the high surface 
brightnesses (evolution $>$ 3~mag) which 
may be detected at high redshift with a good S/N, it would not be possible to get reliable 
estimates of their morphological parameters $A$ and $C_A$ and therefore to classify
them at z$>$2.
In Fig.~\ref{logrhlIab} we compare the detection limit (with an exposure time
of 2.5~ksec) reached by Roche et al.
(\cite{Roche96}) in the ($Log (r_{hl})$ vs.  $I_{AB}$) diagram for I-band
exposures to our measured values (in the I-band as well).  It was necessary to
 change
Roche et al.'s (\cite{Roche96}) Johnson I magnitudes to the AB systems by
applying the relation $I_{AB} = I + 0.52$. Our spiral galaxies could have
been detected by Roche et al. (\cite{Roche96}) assuming boosts $\ge$ 1-2~mag
for M51, M100 and M101 but boosts $\ge$ 3-4~mag for M33 and M81. The size of M101
is in the upper bin in the $r_{hl}$ distribution presented by Roche et al. 
(\cite{Roche98}). The other galaxies have $r_{hl}$ in the observed range. 

Note that our simulations are more optimistic that actual HST observations (see Sect.~3),
and it would be even more difficult to detect them. However, the detection is not the whole story
and an additional caveat appears. We have only been able to quantitatively 
estimate the asymmetries and concentration for M51, M100 and M101 with  
large boosts. This means that we would not be able to classify those
galaxies unless very large boosts were applied to the brightest galaxies. 
Moreover, as we will see in the next section, these galaxies would not appear as
spirals anyway.

Models of galaxy formation in hierarchical cold dark matter (CDM) cosmogonies
predict that Milky Way-like disks cannot form at $z > 1$ in a universe with
$\Omega_0 \sim 1$ while lower constraints come from a low-$\Omega_0$
universe (Mo et al.  \cite{Mo98}).  Note, however, that these scenarios predict
that early disks may be present at high redshifts but with a size significantly
smaller than the disks observed today.  From an analysis of the NTT Deep Field,
Poli et al.  (\cite{Poli99}) found that the size distribution of a sample of
disk-dominated galaxies peaks at very small sizes, $r_{hl} \approx 2.5$ kpc,
corresponding in their sample to $r_{hl} \approx$ 0.3 $arcsec$.  This is also
consistent with the HST Medium Deep Survey (MDS) results (Roche et al.  \cite{Roche98}).
Comparing their results to the CDM predictions, Poli et al.  (\cite{Poli99})
show a general agreement but noticed a possible excess of faint, small-sized
galaxies.  Our simulations show that we could not draw any definite
conclusions on the existence of large spiral disks at redshifts $z > 1$ as
stated by Mao, Mo and White (\cite{Mao98}).  Actually, the observational
constraints from the fact that we could not detect them with 
the {\it HST} and NTT are too small.

What are the disk-dominated galaxies seen by Roche et al.  (\cite{Roche98}) at z
$\approx$ 3-4 and Poli et al.  (\cite{Poli99}) at I $\le$ 25~?  Our simulations
show that we do not expect any variations of the size with the redshift due to dimming
of surface brightness.  Roche et al.  (\cite{Roche98}) conclude that the
$r_{hl}$ distributions are in agreement with a size-luminosity evolution model
where spiral galaxies undergo a small size evolution below z $\le$ 1.5 but are
smaller by a factor of the order of 2 at z $\approx$ 3.  In addition, Mo et al.
(\cite{Mo98}) models for the formation of galactic disks would tend in the same
direction.  It is therefore a subject that deserves further work and we will
analyze, in a follow-up paper, scenarios where we vary the size of the disk
assuming, for instance, a radial variation of the star formation history as in
Roche et al.  (\cite{Roche96}).

\section{Morphology of disks at high redshifts}

\begin{figure} \resizebox{10cm}{18cm}{\includegraphics{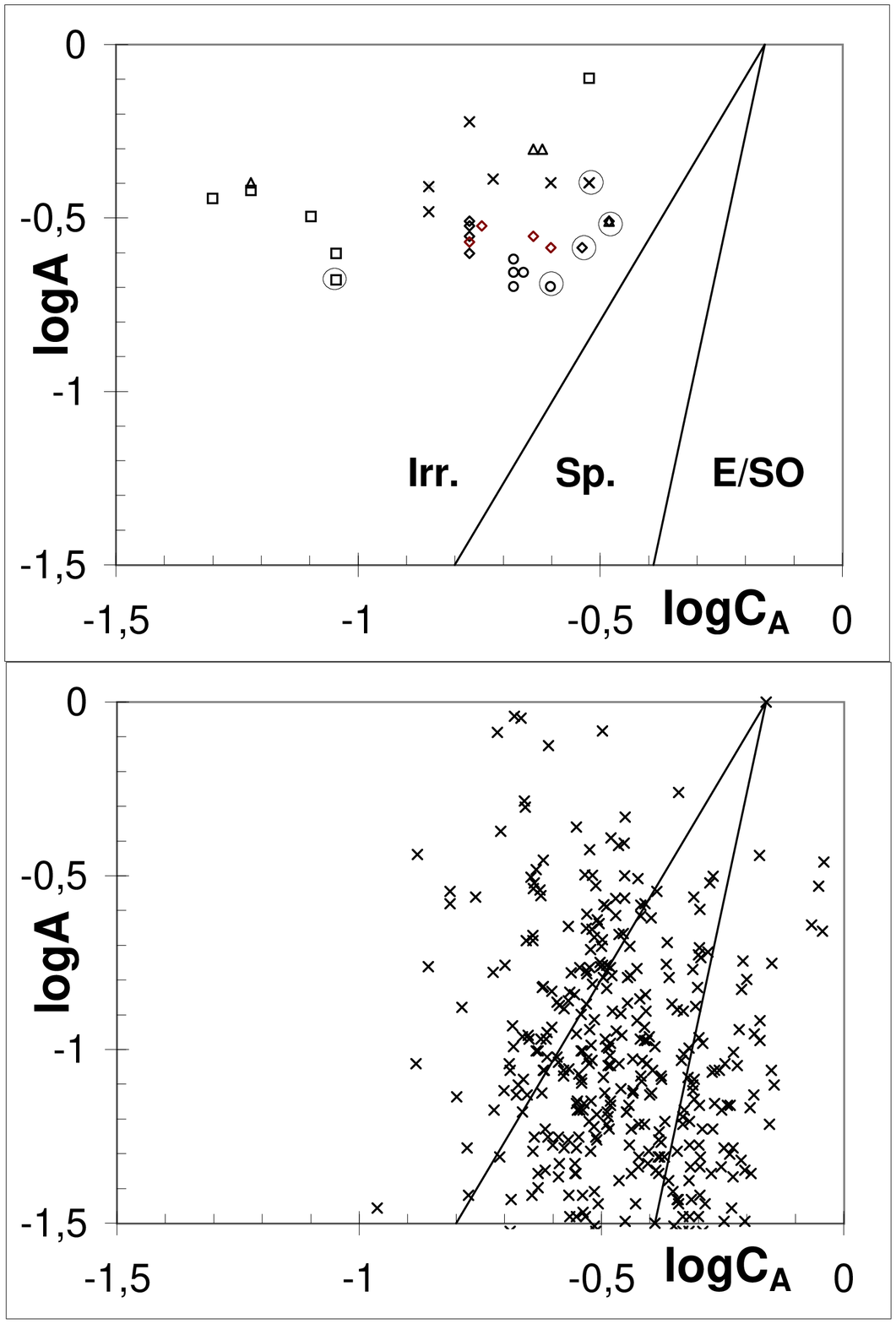}}
\caption[]{Upper panel~: the distribution of our sample of 
rest-frame
and artificially redshifted galaxies in the log $A$ versus log $C_A$
asymmetry-concentration diagram.  The redshifted galaxies are all lying in the
irregular area as defined at visible wavelengths by Abraham et al.  
(\cite{Abraham96b}).  An important
result, however, is that rest-frame ultraviolet spiral galaxies are found in
this area
as well.  This confirms the important role of the observation wavelength range.
On this basis, we should therefore expect no or little morphology evolution 
of the spiral galaxies with the redshift.
The lower panel shows as a comparison the classification of galaxies in the
{\it} HHDF (Abraham et al.  \cite{Abraham96b}). The symbols are
defined as follows~: triangle for M33, diamond for M51, box for M81, circle for
M100 and cross for M101. Symbols within a larger circle represent galaxies at 
z~$\approx$~0.}
\label{logAlogC} \end{figure}

Using our own sample of local spiral galaxies observed in UV with FOCA, we
 simulate high-redshift observations and  we also estimate
asymetries and concentration for our simulated galaxies.  Tabs.  3-7 and
Fig.~\ref{logAlogC} present our results.  Fig.~\ref{logAlogC} is the (Log~$A$ vs.
log~$C_A$) diagram where the dotted lines represent the separations of Hubble types
reported by Abraham et al.  (\cite{Abraham96b}).  As noted above, we have assumed
several scenarios for the luminosity evolution.  The first point to note is that
all the galaxies fall in the top-left area corresponding to the irr/pec/mrg
galaxies.  We do confirm the previous qualitative results that spiral galaxies
observed in a UV rest-frame appear more irregular.  Indeed, this effect is
clearly present even at z $\approx$ 0 but it must be pointed out the the morphology is 
very stable and does not change with the redshift. From z=0 to the highest
explored redshifts, we observed similar concentrations and 
asymmetries for a given galaxy.  The migration of spiral galaxies towards more
asymetrical areas is mainly caused by the clumpiness of the star formation
regions observed in UV.  The symbols corresponding to the redshifted galaxies in
the {\it HST} filters fall very close to their z $\approx$ 0 parent galaxy.  
Galaxies where the
evolution is assumed to be proportional to e$^{- t / \tau}$ fall in the 
diagram at -1.4 $\le$log~$C_A$ $\le$ -0.5. This quite large range is in fact 
due to M81 (log~$C_A \le$ -1.0) whereas the other galaxies show similar 
concentrations (-0.9 $\le$log~$C_A$ $\le$ -0.5). Moreover, note that the
concentration is not consistent with the usual values measured in the visible.
Indeed, M81 is the most early-type galaxy but appears as the 
least concentrated galaxy in Fig.~\ref{logAlogC}. Clearly, more UV templates
must be studied to test concentration as a morphological discriminator between
early and late type galaxies. All spiral galaxies lie in 
a very narrow log A range (-0.7 $\le$ log A $\le$ -0.2). They are very 
asymmetric compared
to their optical morphology and are located in the irregular domain.  Furthermore,
the degeneracy of the log $A$ parameter in UV is very limiting for
morphology studies. Other ways of measuring the asymmetry have been studied. 
For instance, Rudnick \& Rix (\cite{Rudnick98}) use the Fourier amplitudes of 
the image.
Kornreich et al. (\cite{Kornreich98}) compare the relative fluxes of 
trapezoidal areas distributed around the center of the galaxy.
Even for  scenarios where the galaxy is uniformely boosted by a
magnitude ranging from 0 $\le$ $\Delta m$ $\le$ 4, the increase of S/N does not
change our conclusion and the galaxies remain in the same area whatever the
scenario. This large asymmetry - low concentration morphology is intrinsic
to the UV.

\section{Conclusion}

The main results of this paper are summarized as follows~:

\begin{itemize} 
\item As expected, galaxies observed in UV are very different
from their visible counterpart as the young populations are predominent.  These
galaxies appear more clumpy.  
\item Local galaxies with large disk projected at
high redshift (0.65 $\le$ z $\le$ 2.94) would be hardly detected with the {\it
HST} in 10-hr exposure times.  
Consequently, no strong observational constraints can be set on the presence or
absence of large galaxy disks at high z.  
\item A quantitative analysis of the
UV morphology of disk galaxies shows that there is a clear trend for spirals to
move towards more irregular morphology types (morphological {\sl K}-correction).
If the concentration parameter appears to be a 
good discriminator between early and late type galaxies, the asymmetry is degenerate
 and all the galaxies have very
similar asymmetry values.  New ways of measuring the asymmetry need to be
explored.  
\item The location of a given galaxy in the (log~$A$ vs.  log~$C_A$)
diagram is very stable and does not depend on the redshift or the S/N as soon as
the rest-frame wavelength is in UV.  Therefore, rest-frame visible imaging is
not helpful and it would be useful to define a multi-wavelength morphology
system.  \end{itemize}

This paper stresses the need to keep on working on the UV morphology of galaxies
if we wish to be able to understand the objects observed at high
redshift. It is therefore necessary to define new tools for classifying galaxies.
A number of objects have been observed by FOCA, UIT-Astro-2 and other
UV imagers. However, the database is still too small for a significative
study. Hopefully, GALEX, the GALaxy Evolution eXplorer will complete a UV survey of
the sky within the next years and the accumulated data will be crucial to
progress in the understanding of the formation and evolution of galaxies.

\begin{acknowledgements} We thank J.-M.  Deltorn and A.  Boselli for helpful and
stimulating discussions about this work.  The FOCA balloon program has been
conducted jointly by the Laboratoire d'Astronomie Spatiale (now Laboratoire
d'Astrophysique de Marseille) and the Observatoire de Gen\`eve.  Financial
support was provided by the Centre National d'Etudes Spatiales (CNES) and
the Fonds National de la Recherche Suisse (FNRS).

\end{acknowledgements}

\end{document}